\documentclass[12pt,a4paper]{article}
\ifcase0
\or\documentclass[12pt,a4paper]{article}
\or\documentclass[11pt,a5paper]{article}
\or\documentclass[12pt,a4paper]{refart}
\settextfraction {0.8}
\fi

\usepackage{euler,amsmath,defns,url,graphicx,color}
\IfFileExists{ajr.sty}{\usepackage{ajr}
    \hypersetup{pdfauthor={A. J. Roberts},
    pdftitle={Model dynamics on a multigrid across multiple length and time scales}, 
    pdfcreator={Tony Roberts, \today},
    pdfproducer={Dept Maths and Computing, USQ}
    }}{}
\allowdisplaybreaks

\newcommand{\ded}{\bar\delta}
\newcommand{\mud}{\bar\mu\ded}
\newcommand{\Ed}{\bar E}
\newcommand{\cM}{{\cal M}}
\newcommand{\op}[1]{{\cal #1}}
\newcommand{\cop}[1]{\bar{\op{#1}}}
\newcommand{\g}[1]{^{(#1)}}
\newcommand{\cT}{{\cal T}}

\newcommand{\gammb}{\bar\gamma}
\newcommand{\cc}{{\mathfrak C}}

\title{Model dynamics on a multigrid across multiple length and time scales} 

\author{A.~J. Roberts\thanks{Computational Engineering and Sciences Research Centre, Dept Maths and Computing, University of Southern Queensland, Toowoomba, Queensland~4350, Australia.}}

\begin{document}

\maketitle

\begin{abstract}
Most methods for modelling dynamics posit just two time scales: a fast and a slow scale.  But many applications, including many in continuum mechanics, possess a wide variety of space-time scales; often they possess a continuum of space-time scales.  I discuss an approach to modelling the discretised dynamics of advection and diffusion with rigorous support for changing the resolved spatial grid scale by just a factor of two. The mapping of dynamics from a finer grid to a coarser grid is then iterated to generate a hierarchy of models across a wide range of space-time scales, all with rigorous support across the whole hierarchy.  This approach empowers us with great flexibility in modelling complex dynamics over multiple scales.
\end{abstract}

\tableofcontents

\section{Introduction}
\label{sec:1}

Multiscale methods promise efficient computation and simulation of many important physical systems~\cite[e.g.]{Brandt01}. Dolbow et al.~\cite{Dolbow04} identify that critical applications include fuel cells, subsurface contaminant transport, protein folding, climate simulations, and general networks.  
Most multiscale mathematical methods for modelling disparate scales presume just two scales: small lengths and large lengths; fast times and long times; fast variations and slow variations; microscopic and macroscopic~\cite[e.g.]{Dolbow04, E04}.  Most such methods then seek \emph{effective} models or properties on the large\slash long\slash slow macroscales by `averaging\slash homogenising over' the small\slash fast microscales~\cite[e.g.]{E04, Pavliotis06a}.
Here we explore a novel mathematical framework to modelling dynamics over many length and time scales; the framework is supported by modern dynamical systems theory.

Of course most two scale modelling methods will also work over many scales.  The same techniques that construct and support a slow model of rapid variations may also apply to construct and support a superslow model of the slow variations.  The same techniques may then also apply to construct and support a megaslow model of the superslow variations; and so on.  Crucially, in most established methods each of these constructions require a large `spectral gap'; that is, they require an absolutely clear separation between the micro and the macro scales; a parameter such as~$\epsilon$ measures the scale separation, and the requirement for extreme scale separation is provided by theorems invoking ``as $\epsilon\to0$''.
In contrast, multigrid iteration for solving linear equations transforms between length scales that are different by (usually) a factor of two~\cite[e.g.]{Briggs01}; some variants of multigrid iteration use an even smaller ratio of length scales~\cite[e.g.]{Roberts99d}.  Recently Brandt~\cite{Brandt06} proposed a method for molecular dynamics without large scale separation using so-called systematic upscaling.  Analogously, here we explore modelling \emph{dynamics} on a hierachy of length scales that differ by a factor of two and hence the `spectral gap' is finite and typically much smaller than required by popular extant methods for modelling dynamics.  Section~\ref{sec:two} rigorously supports such models with centre manifold theory~\cite[e.g.]{Carr81}.

Recall that multigrid iteration, using restriction and prolongation operators, transforms between length scales differing by a factor of two~\cite[e.g.]{Briggs01}.  The full multigrid iteration involves iterating the restriction and prolongation transformations to cross large changes in length scales by taking many coarsening or refining transforms where each step changes the resolved length scales by a factor of two.  Similarly, Section~\ref{sec:mad} starts our modelling of dynamics by exploring a transformation of dynamics from one length scale to another that is a factor of two coarser.
Section~\ref{sec:mad}, see~\eqref{eq:gaddif}, demonstrates that to some controlled approximation the discrete advection diffusion equation
\begin{equation}
    \frac{du_j}{dt}= -c\rat12(u_{j+1}-u_{j-1}) 
    + d(u_{j+1}-2u_j+u_{j-1})\,,
    \label{eq:addif}
\end{equation}
for evolving grid values~$u_j(t)$ on a grid of spacing~$h$ (and hence with `advection speed'~$ch$ and `diffusion'~$ d h^2$) is, on the coarser grid of spacing ${\bar h}=2h$\,, justifiably modelled by
\begin{eqnarray}
    &&\frac{d{\bar u}_j}{dt}\approx -\bar c\rat12({\bar u}_{j+1}-{\bar u}_{j-1}) 
    +\bar d({\bar u}_{j+1}-2{\bar u}_j+{\bar u}_{j-1})
    \nonumber\\\text{where}&&
    \bar c=\rat12c
    \qtq{and}
    \bar d=\rat14 d+\frac{c^2}{16 d}\,,
    \label{eq:addif2}
\end{eqnarray}
for evolving coarse grid values~${\bar u}_j$; these coarse grid values are defined to be the fine grid values at every second point on the fine grid, ${\bar u}_j=u_{2j}$\,.   Intriguingly, the key to the approach is to take one step backwards in order to take two steps forward:  at any level we embed the dynamics~\eqref{eq:addif} in a higher dimensional problem, then analysis systematically derives the lower dimensional, macroscale model~\eqref{eq:addif2}.  The geometric approach to modelling of both centre manifold theory~\cite[e.g.]{Carr81} and normal form theory~\cite[e.g.]{Murdock03} justifies the model~\eqref{eq:addif2} using a similar approach to that of holistic discretisation~\cite[e.g.]{Roberts98a}.  The enhancement of the diffusion by~$c^2/(16 d)$ evident in~\eqref{eq:addif2} on the coarse grid  comes from resolving the dynamics on the finer grid in constructing the model on the coarser grid:\footnote{Increasing the dissipation at coarser levels has proven effective in multiscale methods for compressible fluid flows~\cite[p.10, e.g.]{Brandt01}.} the enhancement ensures the coarse model~\eqref{eq:addif2} stably models the fine grid dynamics~\eqref{eq:addif}; intriguingly this enhanced dispersion is \emph{precisely} that implicit in cyclic reduction, a multigrid method, to find an equilibrium of such advection-dispersion problems, but here derived for dynamic problems with a different theoretical base.
The coarse model~\eqref{eq:addif2} implicitly prescribes a `restriction operator' that transforms the dynamics of advection-dispersion from one grid to another with twice the spacing.

\begin{figure}
\centering
\setlength\unitlength{2ex}
\newcommand{\blob}{\circle*{0.5}}
\begin{picture}(33,15)(0,0)
\thicklines
\put(-1.5,1){\vector(0,1){7}}
\put(-2,9){\rotatebox{90}{level $\ell$}}
\put(0,0){\vector(1,0){15}\ $x$\ space}
\multiput(0,1.5)(2,0){17}{\blob}
\put(0,2){$u_0\g0$}
\put(2,2){$u_{1}\g0$}
\put(4,2){$u_{2}\g0$}
\put(6,2){$u_{3}\g0$}
\put(8,2){$u_{4}\g0$}
\put(10,2){$u_{5}\g0$}
\put(12,2){$u_{6}\g0$}
\put(14,2){$u_{7}\g0$}
\put(16,2){$u_{8}\g0$}
\put(18,2){$u_{9}\g0$}
\put(20,2){$u_{10}\g0$}
\put(22,2){$u_{11}\g0$}
\put(24,2){$u_{12}\g0$}
\put(26,2){$u_{13}\g0$}
\put(28,2){$u_{14}\g0$}
\put(30,2){$u_{15}\g0$}
\put(32,2){$u_{16}\g0$}
\multiput(0,4.5)(4,0){9}{\blob}
\put(0,5){$u_0\g1$}
\put(4,5){$u_1\g1$}
\put(8,5){$u_2\g1$}
\put(12,5){$u_3\g1$}
\put(16,5){$u_4\g1$}
\put(20,5){$u_5\g1$}
\put(24,5){$u_6\g1$}
\put(28,5){$u_7\g1$}
\put(32,5){$u_{8}\g1$}
\multiput(0,7.5)(8,0){5}{\blob}
\put(0,8){$u_0\g2$}
\put(8,8){$u_1\g2$}
\put(16,8){$u_2\g2$}
\put(24,8){$u_3\g2$}
\put(32,8){$u_{4}\g2$}
\multiput(0,10.5)(16,0){3}{\blob}
\put(0,11){$u_0\g3$}
\put(16,11){$u_1\g3$}
\put(32,11){$u_2\g3$}
\multiput(0,13.5)(32,0){2}{\blob}
\put(0,14){$u_0\g4$}
\put(32,14){$u_1\g4$}
\end{picture}
\caption{schematic picture of the multigrid underlying the multiscale description of the dynamics: as usual for multigrids, the grids of spacing $h\g\ell=2^\ell h$ are stacked vertically with relatively coarse grids above fine grids; the dynamic variables of the hierarchy of discrete models are~$u_j\g\ell$.}
\label{fig:hsg}
\end{figure}
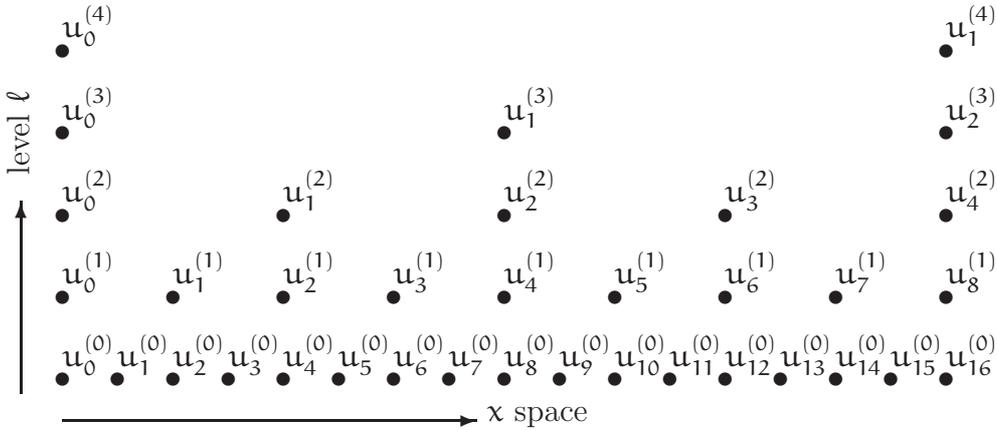

Others also explore dynamics across space-time scales. 
Griebel, Oeltz \& Vassilevski~\cite{Griebel06} developed  space-time multigrid numerics to find optimal control of problems governed by parabolic differential equations.  They base their cross-scale transformation on an algebraic multigrid.  The systematic upscaling by Brandt~\cite{Brandt06} uses multigrid ideas to progressively coarsen atomic simulations of polymer folding.  These approaches are largely computational whereas here we develop algebraic transformations that then are used computationally.  Another major difference is that the slow manifolds constructed here provides a coarsening and interpolation, across length scales, that is specifically adapted to the dynamics of the problem rather than being imposed on the problem.  Section~\ref{sec:caasm} constructs the slow manifolds  by systematically approximating exact closures provided by the fine scale dynamics.

Section~\ref{sec:multi} explores iterating our transformation to model dynamics across each and every intervening length scale.  For example,
repeating the transformation from fine~\eqref{eq:addif} to coarse~\eqref{eq:addif2}
gives a hierarchy of models all of the form of the advection-dispersion equation~\eqref{eq:addif} but with differing coefficients.  At the $\ell$th~level, with grid spacing $h\g{\ell}=2^\ell h$\,, the corresponding grid values~$u_j\g{\ell}$ evolve according to~\eqref{eq:addif} but with coefficients $c\g{\ell}$~and~$ d\g{\ell}$ determined by the recurrence
\begin{equation}
    c\g{\ell +1}=\rat12c\g{\ell}
    \quad\text{and}\quad
     d\g{\ell +1}=\rat14 d\g{\ell}+\frac{{c\g{\ell}}^2}{16 d\g{\ell}}\,.
    \label{eq:addifn}
\end{equation}
On successively coarser grids the coefficients thus are
\begin{equation}
    c\g{\ell}=\frac c{2^\ell}
    \qtq{and}
     d\g{\ell}=\frac {|c|}{2^{\ell +1}}\tilde d\g{\ell}
    \qtq{where}
    \tilde d\g{\ell +1}=\frac12\left(\tilde d\g{\ell}
    +\frac1{\tilde d\g{\ell}}\right)\,.
    \label{eq:cgcta}
\end{equation}
Observe that $\tilde d\g{\ell}\to1$  quickly as $\ell $~increases (as~\eqref{eq:cgcta} is equivalent to Newton's iteration to find the zeros of $\tilde d^2-1$); hence as the grid coarsens, the $\ell $th~level model quickly becomes simply the upwind model
\begin{displaymath}
    \frac{du\g{\ell}_j}{dt}\approx-c\frac{u\g{\ell}_j-u\g{\ell}_{j-1}}{2^\ell}
    \quad\text{when }c>0\,.
\end{displaymath}
Our multigrid modelling transformation naturally recognises that advection dominates diffusion on coarse grids: the cross scale transformation, the map from fine~\eqref{eq:addif} to coarse~\eqref{eq:addif2} as summarised by~\eqref{eq:addifn}, not only preserves the advection speed, but also models the advection in a stable scheme that preserves non-negativity.  Further, in the absence of advection, $c=0$\,, the transformation~\eqref{eq:addifn} preserves the effective diffusion across all scales: $ d\g{\ell +1}=\rat14 d\g{\ell}$\,.  These are some simple results.
Section~\ref{sec:multi} explores further issues in transforming both linear and nonlinear discrete dynamics across many scales.

The centre manifold and normal form~\cite[e.g.]{Elphick87a, Cox93b,
Murdock03} approach established here provides a framework for dynamical modelling that links what are conventionally called multigrid~\cite[e.g.]{Briggs01}, wavelets~\cite[e.g.]{Daubechies92}, multiple scales~\cite[e.g.]{Pavliotis07}, and singular perturbations~\cite[e.g.]{Verhulst05}.  This framework applies to not only the linear dynamical systems that are the main focus of this article, but also applies to nonlinear systems~\cite[e.g.]{Roberts98a, MacKenzie05a} and to stochastic systems~\cite[e.g.]{Chao95, Arnold03, Roberts03c, Roberts06k}.  Here, because it is simplest, we focus on transforming dynamics within the same algebraic form, but in principle the methodology can support the emergence, via nonlinear interactions, of qualitatively different dynamics on macroscales (as promoted by the heterogeneous multiscale method~\cite[e.g.]{E04}).   By rationally transforming across both space and time scales, a long term aim of this approach is to empower efficient simulation and analysis of multiscale systems at whatever level of detail is required and to a controllable error.

This approach to transformation from one scale to anther may in the future illuminate complex systems simulations on both lattices and with cellular automata.

\section{Centre manifold theory supports multiscale models}
\label{sec:two}

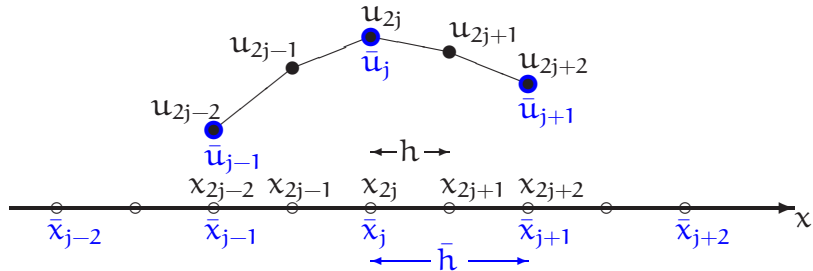
\begin{figure}
\newcommand{\B}{\color{blue}}
\centering
    \setlength{\unitlength}{0.1em}
    \begin{picture}(260,70)
      \put(0,-10){
          \thinlines   
	  \put(65,50){
	  \put(0,0){\line(5,4){25}}
	  \put(25,20){\line(5,2){25}}
	  \put(50,30){\line(5,-1){25}}
	  \put(75,25){\line(5,-2){25}}
	  \put(0,0){\B\circle*{6}} 
	  \put(0,0){\circle*{3}} 
          \put(-3,-11){\B${\bar u}_{j-1}$}
          \put(-20,5){$u_{2j-2}$}
      \put(25,20){\circle*{4}}
          \put(5,25){$u_{2j-1}$}
	  \put(50,30){\B\circle*{6}} 
	  \put(50,30){\circle*{3}} 
          \put(47,19){\B${\bar u}_j$}
          \put(47,35){$u_{2j}$}
      \put(75,25){\circle*{4}}
          \put(75,30){$u_{2j+1}$}
	  \put(100,15){\B\circle*{6}} 
	  \put(100,15){\circle*{3}} 
          \put(97,4){\B${\bar u}_{j+1}$}
          \put(97,20){$u_{2j+2}$}
	  }
	  \put(124,40){$h$}
      \put(132,43){\vector(1,0){8}}
      \put(123,43){\vector(-1,0){8}}
	  \put(136,5){\B${\bar h}$}
      \put(146,8){\B\vector(1,0){19}}
      \put(134,8){\B\vector(-1,0){19}}
      \put(12,15){\B${\bar x}_{j-2}$}
      \put(62,15){\B${\bar x}_{j-1}$}
      \put(112,15){\B${\bar x}_j$}
      \put(162,15){\B${\bar x}_{j+1}$}
      \put(212,15){\B${\bar x}_{j+2}$}
      \put( 57,30){$x_{2j-2}$}
      \put( 82,30){$x_{2j-1}$}
      \put(112,30){$x_{2j}$}
      \put(137,30){$x_{2j+1}$}
      \put(162,30){$x_{2j+2}$}
	  \multiput(15,25)(25,0){9}{\circle{3}}
	  \thicklines
	  \put(250,19){$x$}
	  \put(0,25){\vector(1,0){250}}
      }
    \end{picture}
\caption{schematic picture of the equi-spaced fine grid,~$x_j$, with spacing~$h$, and grid values~$u_j$\,.  The coarse grid,~${\bar x}_j$, with spacing ${\bar h}=2h$\,, and grid values~${\bar u}_j$ is superposed.}
\label{fig:grid}
\end{figure}%
            
This section establishes new theoretical support for coarsening dynamics from a fine grid to a coarse grid of twice the spacing.  Suppose the fine grid has grid points~$x_j$, spacing~$h$ as shown in Figure~\ref{fig:grid}, and has grid values~$u_j(t)$ evolving in time.  The figure also shows the coarse grid points~${\bar x}_j=x_{2j}$\,, spacing ${\bar h}=2h$\,, and the definition of the evolving coarse grid values\footnote{This is the amplitude condition that ensures the relation between the coarse grid values and the fine grid values are unique and well defined.  Others, such as Brandt~\cite{Brandt06}, prefer using averages of the fine grid values as coarse variables which would also be acceptable.  Centre manifold theory allows the coarse variables to be \emph{any} reasonable measure of the amplitude of the fine grid variables.}
\begin{equation}
    {\bar u}_j(t)=u_{2j}(t)\,.
    \label{eq:uamp}
\end{equation}
Mostly, an overbar denotes variables and operators on the coarser grid, and unadorned variables are those on the finer grid.  Using overdots to denote time derivatives, the question is: 
how do we transform the evolution $\dot u_j=\op L u_j$\,, for some fine scale local operator~$\op L$, to a coarse evolution $\dot {\bar u}_j=\cop L {\bar u}_j$ on the coarse grid? 

The theoretical support for multiscale modelling outlined by this section applies equally well to nonlinear dynamics:  
Section~\ref{sec:anbd} briefly explores the specific nonlinear  advection-dispersion of a discrete Burgers' equation.  Assume the fine spatial grid is periodic with $m$~grid points: that is, for definiteness assume the grid is periodic in space~$x$ with period~$mh$.  For conciseness, write equations in terms of centred mean~$\mu$ and difference~$\delta$ operators
\cite[Ch.~7, e.g.]{npl61} acting on the fine grid.  Thus the advection-dispersion equation~\eqref{eq:addif}, but now including some `nonlinearity'~$f_j$ with some parameters~$\vec\epsilon$, is
\begin{equation}
    \dot u_j
     = \big\{ -c\mu\delta + d\delta^2 \big\}u_j +f_j(\vec u,\vec \epsilon)\,.
    \label{eq:ad}
\end{equation}
I give three illustrative examples of such nonlinearity: a local reaction could be prescribed by $f_j=\epsilon u_j-u_j^3$\,; a nonlinear advection by $f_j=u_j\mu\delta u_j/(2h)$ as in the discrete Burgers' equation~\eqref{eq:burg0}; whereas linear diffusion in a random medium could be encompassed by $f_j=\epsilon \delta(\kappa_j\delta u_j)/h^2$ for some stochastic diffusivities~$\kappa_j$.

\paragraph{Centre manifold support} We now describe how to support and construct the model on the coarse grid of the fine scale, nonlinearly modified, advection-dispersion dynamics.  

\begin{figure}
\newcommand{\B}{\color{blue}}
\centering
    \setlength{\unitlength}{0.125em}
    \begin{picture}(260,100)
    \put(0,50){
	  \put(124,40){$h$}
      \put(132,42){\vector(1,0){8}}
      \put(123,42){\vector(-1,0){8}}
      \put(50,-35){
	      \put(86,41){\B${\bar h}$}
          \put(96,43){\B\vector(1,0){19}}
          \put(84,43){\B\vector(-1,0){19}}}
      \put(12,15){\B${\bar x}_{j-2}$}
      \put(62,15){\B${\bar x}_{j-1}$}
      \put(112,15){\B${\bar x}_j$}
      \put(162,15){\B${\bar x}_{j+1}$}
      \put(212,15){\B${\bar x}_{j+2}$}
      \put( 7,30){$x_{2j-4}$}
      \put( 32,30){$x_{2j-3}$}
      \put( 57,30){$x_{2j-2}$}
      \put( 82,30){$x_{2j-1}$}
      \put(112,30){$x_{2j}$}
      \put(137,30){$x_{2j+1}$}
      \put(162,30){$x_{2j+2}$}
      \put(187,30){$x_{2j+3}$}
      \put(212,30){$x_{2j+4}$}
	  \multiput(15,25)(25,0){9}{\circle{3}}
	  \thicklines
	  \put(250,19){$x$}
	  \put(0,25){\vector(1,0){250}}
	 }
	 \put(15,45){
	     \thicklines
	     \put(0,0){\B\line(1,0){100}}
	     \multiput(0,0)(25,0){5}{\circle{3}}
	     \put(-9,5){
		     \put(0,0){$v_{j-1,-2}$}
		     \put(25,0){$v_{j-1,-1}$}
		     \put(50,0){$v_{j-1,0}$}
		     \put(75,0){$v_{j-1,1}$}
		     \put(100,0){$v_{j-1,2}$}
	     }
	 }
	 \put(65,25){
	     \thicklines
	     \put(0,0){\B\line(1,0){100}}
	     \multiput(0,0)(25,0){5}{\circle{3}}
	     \put(-5,5){
		     \put(0,0){$v_{j,-2}$}
		     \put(25,0){$v_{j,-1}$}
		     \put(50,0){$v_{j,0}$}
		     \put(75,0){$v_{j,1}$}
		     \put(100,0){$v_{j,2}$}
	     }
	 }
	 \put(115,5){
	     \thicklines
	     \put(0,0){\B\line(1,0){100}}
	     \multiput(0,0)(25,0){5}{\circle{3}}
	     \put(-9,5){
		     \put(0,0){$v_{j+1,-2}$}
		     \put(25,0){$v_{j+1,-1}$}
		     \put(50,0){$v_{j+1,0}$}
		     \put(75,0){$v_{j+1,1}$}
		     \put(100,0){$v_{j+1,2}$}
	     }
	 }
	 \put(180,45){$\longleftarrow(j-1)$th element}
	 \put(200,25){$\longleftarrow j$th element}
	 \put(0,5){$(j+1)$th element $\longrightarrow$}
    \end{picture}
\caption{schematic picture of the equi-spaced fine grid,~$x_j$, with spacing~$h$, and the coarse grid,~${\bar x}_j$, with spacing ${\bar h}=2h$\,.  Three finite elements of the coarse grid are shown in exploded view to illustrate their overlap with neighbouring elements.  Within each finite element new variables~$v_{j,i}(t)$, $-2\leq i\leq 2$\,, replace the fine grid variables~$u_j(t)$.}
\label{fig:explod}
\end{figure}
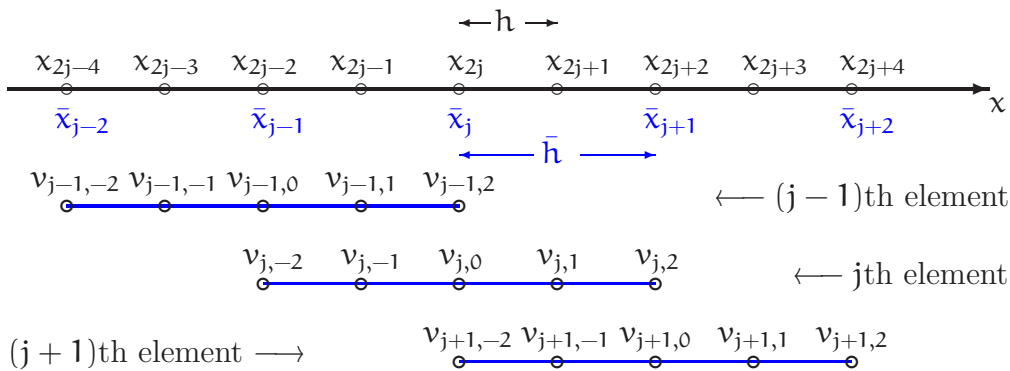%
 
Analogous to holistic discretisation of \pde{}s~\cite[e.g.]{Roberts98a, MacKenzie05a}, divide the $m$-periodic fine grid into
$m/2$~overlapping elements.  Notionally let the $j$th~coarse element stretch from $x_{2j-2}$~to~$x_{2j+2}$ as shown `exploded' in Figure~\ref{fig:explod}.
As shown, denote the evolving fine grid field in the $j$th~element as the 5-tuple $\vec v_j=(v_{j,-2},v_{j,-1},v_{j,0},v_{j,1},v_{j,2})$\,, so that at this stage we have just renamed the fine grid variables,
$u_{2j+i}=v_{j,i}$\,.  Note that the elements overlap: the fine grid values
$u_{2j\pm1}=v_{j,\pm1}=v_{j\pm 1,\mp1}$\,; this overlap empowers us to couple the dynamics in neighbouring elements to derive consistent models as similarly derived for holistic discretisation~\cite{Roberts00a}.\footnote{The overlapping elements may be analogous to the `border regions' of the heterogeneous multiscale method~\cite[e.g.]{E04} and to the `buffers' of the gap-tooth scheme~\cite[e.g.]{Samaey03b}.}  The interelement coupling conditions~\eqref{eq:ibc} determine the fine grid values $v_{j,\pm2}$, at the extremes of each element, and so these are not extra dynamic variables.  But, importantly, consider the overlapping fine grid values $v_{j,\pm1}$~and~$v_{j\pm 1,\mp1}$ as independent dynamic variables satisfying the fine scale discrete equation~\eqref{eq:ad}, namely
\begin{equation}
    \dot v_{j,i}
     = \big\{ -c\mu\delta + d\delta^2 \big\}v_{j,i} +f_j(\vec v_j,\vec \epsilon)\,,
     \quad i=0,\pm1\,,
    \label{eq:adv}
\end{equation}
where these differences and means operate over the fine grid index~$i$.
In essence I extend the dynamics of the $m$~fine grid variables~$u_j(t)$ by an extra $m/2$~variables.  This is the `one step backwards' referred to in the Introduction: in order to rigorously support the modelling of the $m$-dimensional fine scale dynamics by $m/2$~coarse scale variables, I embed the fine scale system in the $3m/2$-dimensional dynamics of these overlapping elements. Section~\ref{sec:nf} shows how to choose these $m/2$~extra degrees of freedom to make forecasts from any given fine grid scale initial condition.

Also analogous to holistic discretisation of \pde{}s~\cite[e.g.]{Roberts98a, MacKenzie05a}, couple neighbouring elements with the conditions
\begin{equation}
    v_{j,\pm2}=\gammb v_{j\pm1,0}+(1-\gammb)v_{j,0}\,,
    \label{eq:ibc}
\end{equation}
where the coupling parameter~$\gammb$ controls the interaction and information flow between elements: 
\begin{itemize}
\item when $\gammb=1$ the elements are fully coupled and the condition~\eqref{eq:ibc} reduces to the statement that the extrapolation of the $j$th~element field to the neighbouring coarse grid points,~$v_{j,\pm2}$, is identical to the neighbouring coarse grid values~$v_{j\pm1,0}(={\bar u}_{j\pm1})$; 
\item when $\gammb=0$ the elements are completely isolated from each other and thus, \emph{linearly}, the new fine grid values~$v_{j,i}$ evolve quickly to be constant in each element.
\end{itemize}
This equilibrium when $\gammb=0$\,, or space of equilibria depending upon the nonlinearity~$f_j$,  forms the base for the slow manifold model which when evaluated at $\gammb=1$ gives the desired model for the fully coupled dynamics.

I use the overbar in~$\gammb$ because it moderates information flow between the elements forming the coarse grid.  By working to an error~$\Ord{\gammb^n}$ we account for interactions between the dynamics in an element and its $n-1$ neighbours on either side.  Thus we transform \emph{local} dynamics on a fine grid to \emph{local} dynamics on a coarse grid as in other multiscale approaches~\cite[e.g.]{Brandt06}.  The size of the locality depends upon the order of error in the coupling parameter~$\gammb$.

\paragraph{The decoupled dynamics have a useful spectral gap}  Set $\gammb=0$ to decouple the elements, and neglect the nonlinearity by linearisation.  Then, independently of all other elements, the linear dynamics in the $j$th~element are governed by the differential-algebraic system
\begin{equation}
    \begin{bmatrix}
        0\\\dot v_{j,-1}\\\dot v_{j,0}\\\dot v_{j,1}\\0
    \end{bmatrix}=
    \begin{bmatrix}
        1&0&-1&0&0\\
        \rat12c+ d&-2 d&-\rat12c+ d&0&0\\
        0&\rat12c+ d&-2 d&-\rat12c+ d&0\\
        0&0&\rat12c+ d&-2 d&-\rat12c+ d\\
        0&0&-1&0&1
    \end{bmatrix}
    \begin{bmatrix}
        v_{j,-2} \\ v_{j,-1}\\ v_{j,0}\\ v_{j,1}\\ v_{j,2}
    \end{bmatrix}\,.
     \label{eq:fined}
\end{equation}
Seeking solutions proportional to $e^{\lambda t}$ this set of linear 
\ode{}s has three eigenvalues and three corresponding eigenvectors:
\begin{equation}
    \lambda=0,-2 d,-4 d\,;
    \qtq{and}
    \begin{bmatrix}
        1\\1\\1\\1\\1
    \end{bmatrix},\quad
    \begin{bmatrix}
        0\\\rat12c- d\\0\\\rat12c+ d\\0
    \end{bmatrix},\quad
    \begin{bmatrix}
        1\\-1\\1\\-1\\1
    \end{bmatrix}.
    \label{eq:fineev}
\end{equation}
From these, any zig-zag structures within an element decay exponentially quickly, and hence these decoupled dynamics results in constant solutions in each element arising on a time scale of~$1/d$\,.  Over all the $m/2$~decoupled elements these piecewise constant solutions form an $m/2$~dimensional linear subspace of equilibria, the so-called slow subspace, in the $3m/2$~dimensional state space of the fine grid values $v_{j,0}$~and~$v_{j,\pm1}$. Centre manifold theory for deterministic systems~\cite[e.g.]{Carr81, Murdock03} or for stochastic systems~\cite[\S8.4, e.g.]{Boxler89, Arnold03} then assures us of the following three part theorem. 
\begin{theorem}\label{lem:cm}
    For some domain of finite non-zero coupling parameter~$\gammb$, and if nonlinear, some neighbourhood of the origin in $(\vec u,\vec\epsilon)$: 
\begin{enumerate}
    \item \label{it:cm1}  there exists an $m/2$-dimensional, invariant \emph{slow
    manifold}~$\cM$ of the coupled dynamics of the discrete
    nonlinearly perturbed, advection-dis-per\-sion~\eqref{eq:adv} with coupling
    conditions~\eqref{eq:ibc}---with a dimension corresponding to each
    of the $m/2$~coarse grid elements;

	\item \label{it:cm2} the dynamics on the slow manifold~$\cM$ are approached
	exponentially quickly, roughly like $\exp(-2 d t)$, by all initial
	conditions~$v_{j,i}(0)$ of the fine grid values in some finite neighbourhood
	of~$\cM$---that is, the slow manifold dynamics faithfully model for
	long times generic solutions of the coupled dynamics;

	\item \label{it:cm3} we may construct the slow manifold model to some order of error in~$\gammb$, $|\vec\epsilon|$ and $|\vec {\bar u}|$ by solving the governing, nolinear, discrete advection-dispersion~\eqref{eq:ad} with coupling conditions~\eqref{eq:ibc} to residuals of the same order.
	
	Two broad cases arise: if the nonlinearity $f_j=0$ whenever $v_{j,i}$~is independently constant in each element---for example the Burgers'-like nonlinearity $f_j=u_j\mu\delta u_j/(2h)$---then the approximation is global in the coarse grid variables~$\vec {\bar u}$; alternatively, whenever $f_j\neq0$ for $v_{j,i}$~independently constant in each element---for example the reaction $f_j=\epsilon u_j-u_j^3$---then the approximation is local to the origin in~$\vec {\bar u}$ 
\end{enumerate}
\end{theorem}

Like systematic upscaling~\cite[pp.6,9]{Brandt06} and other multiscale methods, this approach uses equilibrium concepts.  But one crucial difference is that centre manifold theory guarantees that the same separation of dynamics occurs in a \emph{finite} neighbourhood about equilibria and hence supports the separation of coarse scale dynamics from the fine scale occurs for nontrivial dynamics.  This approach provides a systematic alternative to the heuristic Fourier or wavelet decompositions for a `local mode analysis'~\cite[\S8]{Brandt01}: here the local modes are determined by the the dynamical system itself through the shape of the slow manifold.

\paragraph{Finite domain}
After constructing an approximate slow manifold model, we evaluate it for coupling parameter $\gammb=1$ to recover a coarse grid model for the fully coupled dynamics on the fine grid.  Is $\gammb=1$ in the `finite neighbourhood' of theoretical support?  It is for the analogous holistic discretisation of the Burgers' \pde~\cite{Roberts98a}. Similarly, Section~\ref{sec:csda} demonstrates that the fully coupled case, $\gammb=1$\,, is indeed within the neighbourhood of theoretical support for the linear ($f_j=0$) dynamics of~\eqref{eq:adv}.

\section{Coarsen linear advection-dispersion}
\label{sec:mad}

Using the theoretical support of centre manifold theory established by the previous section, this section analyses linear advection-dispersion to provide the multiscale modelling results summarised in the Introduction.   

\subsection{Computer algebra approximates the slow manifold}
\label{sec:caasm}

Elementary algebra readily constructs general slow manifold models~\cite[e.g.]{Roberts96a, MacKenzie05a}.  We solve the fine grid, linear, discrete, advection-dispersion equation~\eqref{eq:adv} with coupling conditions~\eqref{eq:ibc} by seeking solutions parametrised by the evolving coarse grid values~${\bar u}_j(t)$: 
\begin{equation}
    v_{j,i}(t)=V_{j,i}(\vec {\bar u},\gammb)
    \qtq{such that}
    \dot {\bar u}_j=G_j(\vec {\bar u},\gammb)\,,
\end{equation}
for some functions $V_{j,i}$~and~$G_j$ to be determined by the iterative algorithm~\cite{Roberts07i}.  The base approximation is the slow subspace of equilibria:
\begin{displaymath}
    v_{j,i}(t)=V_{j,i}(\vec {\bar u},\gammb)\approx {\bar u}_j
    \qtq{such that}
    \dot {\bar u}_j=G_j(\vec {\bar u}_j,\gammb)\approx 0\,.
\end{displaymath}
Computer algebra code~\cite{Roberts07i} systematically refine these slow manifold approximations.  The refining iteration is based upon the residuals of the discrete equation~\eqref{eq:adv} with coupling conditions~\eqref{eq:ibc}.  Centre manifold theory then assures us that the error in approximating the slow manifold model is of the same order in coupling parameter~$\gammb$ as any remaining residual.  For example, to errors~$\Ord{\gammb^2}$, computer algebra~\cite{Roberts07i} constructs the slow manifold~$\cM$ in the $j$th~element as 
\begin{equation}
    \vec V_j=
    \begin{bmatrix}
        (1-\gammb)+\gammb \Ed^{-1} \hfill \\
        1-\rat12\gammb\mud
        +\rat18\gammb(1+\rat{c}{ d}-\rat{c^2}{4 d^2})\ded^2   \\
        1 \hfill \\
        1+\rat12\gammb\mud
        +\rat18\gammb(1-\rat{c}{ d}-\rat{c^2}{4 d^2})\ded^2 \\
        (1-\gammb)+\gammb \Ed \hfill
    \end{bmatrix}{\bar u}_j
    +\Ord{\gammb^2}\,,
    \label{eq:vad}
\end{equation}
in terms of the coarse grid centred difference and mean operators,
$\ded$~and~$\mud$, and shift operator~$\Ed$ (define $\Ed {\bar u}_j={\bar u}_{j+1}$ or equivalently $\Ed=E^2$).  The terms in~\eqref{eq:vad} which are independent of advection~$c$, for the fully coupled $\gammb=1$\,, are classic quadratic interpolation from the surrounding coarse grid values~${\bar u}_j$.  The terms involving advection, flagged by~$c$, arise through accounting for the dynamics of the fine grid values~$u_{2j\pm1}$ and their interaction with the surrounding grid values.  Equation~\eqref{eq:vad} corresponds to the multigrid prolongation operator, but here it is derived by accounting for the fine scale dynamics rather than being imposed.   

The evolution on the slow manifold~\eqref{eq:vad} is then the coarse grid model
\begin{equation}
    \dot {\bar u}_j= \gammb\left[ -\rat12c\mud 
    +\left(\rat14 d+\frac{c^2}{16 d}\right)\ded^2 \right]{\bar u}_j
    +\Ord{\gammb^2}\,.
    \label{eq:gaddif}
\end{equation}
Neglecting the $\Ord{\gammb^2}$ error, evaluate~\eqref{eq:gaddif} at the physically relevant coupling $\gammb=1$ to deduce the coarse grid model~\eqref{eq:addif2} discussed in the Introduction.

\subsection{The coarse scale, slow dynamics model precisely}
Consider further the linear advection-dispersion~\eqref{eq:adv} with interelement coupling conditions~\eqref{eq:ibc}.  The previous subsection constructed an approximation to errors~$\Ord{\gammb^2}$; this subsection gives exact formula for all coupling~$\gammb$.

Seek solutions with structure within the finite elements of the formal operator form $\vec v_j=\exp(t\op K_n)\vec e_n$\,, where, generalising~\eqref{eq:fineev} to non-zero coupling, $\op K_n$~is the $n$th `operator eigenvalue' of the  advection-dispersion \eqref{eq:adv}--\eqref{eq:ibc} and $\vec e_n$~is the corresponding `operator eigenvector'.
Elementary algebra for any coupling~$\gammb$ reveals the three operator eigenvalues are precisely
\begin{equation}
    \op K_{1,3}=2 d\left\{-1 \pm\sqrt{1+\gammb\left[
    \frac14\left(1+\frac{c^2}{4 d^2}\right)\ded^2
    -\frac{c}{2 d}\mud\right] }\right\}
    \qtq{and}
    \op K_2=-2 d\,.
    \label{eq:ops}
\end{equation}
The smallest (least negative) of these, namely~$\op K_1$,  governs the longest time scales in the coupled dynamics.   
For example, the Taylor expansion\footnote{The Taylor expansion converges provided the bracketed coefficient of the coupling~$\gammb$ is of magnitude less than one.  
This convergence occurs provided the differences $\mud$~and~$\ded$ are small enough; 
that is, for solutions varying slowly enough across the grids.} 
in the coupling~$\gammb$ of operator~$\op K_1$ (the plus case above),  upon using the identity $\bar\mu^2 =1 +\rat14\ded^2$\,, agrees with the $\Ord{\gammb^2}$ evolution~\eqref{eq:gaddif}, to the $\Ord{\gammb^3}$ approximation~\eqref{eq:gnpad}, and to the $\Ord{\gammb^5}$ approximation~\eqref{eq:dnlmh} of isotropic dynamics.  
That is, the coarse grid evolution operator $\cop L=\op K_1$\,.

\begin{lemma}
The coarse grid operator~$\cop L=\op K_1$ reproduces exactly the fine grid operator of linear advection-dispersion when the elements are fully coupled.
\end{lemma}
\begin{proof}
In the fully coupled limit, $\gammb=1$\,, the three operator eigenvalues~\eqref{eq:ops} reduce to
\begin{equation}
    \cop L=\op K_1= d(2\bar\mu-2)-\rat12c\ded
    \,,\quad
    \op K_2=-2 d
    \,,\quad
    \op K_3=- d(2\bar\mu+2)+\rat12c\ded\,.
    \label{eq:opeval}
\end{equation}
Relate to the fine grid operators, via the coarse grid shift operator~$\Ed$ and the fine grid shift operator~$E(=\Ed^{1/2})$, by observing
\begin{eqnarray*}&&
    2\bar\mu-2=\Ed^{1/2}+\Ed^{-1/2}-2
    =E+E^{-1}-2=\delta^2
    \\\text{and}&&
    \rat12\ded=\rat12\big(\Ed^{1/2}-\Ed^{-1/2}\big)
    =\rat12\big(E-E^{-1}\big)=\mu\delta\,.
\end{eqnarray*}
Consequently, the coarse grid model 
\begin{displaymath}
    \dot {\bar u}_j=\cop L {\bar u}_j 
    =\big[ d(2\bar\mu-2)-\rat12c\ded\big]{\bar u}_j
    =\big[ d\delta^2-c\mu\delta\big]{\bar u}_j\,,
\end{displaymath} 
is \emph{precisely} the fine grid, linear, advection-dispersion equation~\eqref{eq:ad}, except that, having half the grid points, it does not resolve the fine scale, high wavenumber, spatial structures that the fine grid can resolve.
\end{proof}  

Thus the operator~$\cop L$ does indeed model on the coarse grid all the coarse dynamics inherent in the fine grid advection-dispersion dynamics.  It is only the approximation of~$\cop L$  by a truncated Taylor series, such as in the $\Ord{\gammb^2}$ model~\eqref{eq:gaddif}, that induces errors in the  coarse scale model of the long term dynamics of linear advection-dispersion.

\subsection{Coarse scale dynamics are attractive}
\label{sec:csda}

Consider the spectrum of the advection-dispersion dynamics implicitly described by the operator eigenvalues~\eqref{eq:ops}.  On any regularly spaced grid, the centred mean and difference operators act on Fourier modes as
\begin{displaymath}
    \mu e^{ikj}=\cos(k/2)e^{ikj}
    \qtq{and}
    \delta e^{ikj}=2i\sin(k/2)e^{ikj}
\end{displaymath}
for a component of spatial wavenumber~$k$ relative to the grid spacing; the wavenumber domain is $-\pi<k\leq \pi$\,.  Upon taking the discrete Fourier transform, the operators $\bar\mu$~and~$\ded$ thus transform to $\cos(\bar k/2)=\cos k$ and $2i\sin(\bar k/2)=2i\sin k$\,, respectively, as the fine grid wavenumber $k=\bar k/2$ in terms of the coarse grid wavenumber~$\bar k$.  Thus from~\eqref{eq:ops}  the advection-dispersion dynamics on the fine grid elements has spectrum
\begin{eqnarray}&&
    \lambda_{1,3}=2 d\left\{-1 \pm\sqrt{1+\gammb\left[
    -\left(1+\frac{c^2}{4 d^2}\right)\sin^2\rat{\bar k}2
    -i\frac{c}{2 d}2\sin \rat{\bar k}2\cos \rat{\bar k}2\right] }\right\}
    \nonumber\\&&\text{and}\quad
    \lambda_2=-2 d\,,
    \label{eq:spec}
\end{eqnarray}
for coarse grid wavenumbers $|\bar k|\leq\pi$\,.  Extensive numerical computations strongly suggest that $0\leq\Re{\sqrt{\cdot}}\leq 1$\,, where $\sqrt{\cdot}$ denotes the square root in~\eqref{eq:spec}, for all wavenumbers $|\bar k|\leq\pi$\,, for all coupling $0\leq\gammb\leq1$\,,
and for all advection relative to diffusion,~$c/ d$.  Consequently, the numerics suggest the spectral ordering $\Re\lambda_3 \leq \Re\lambda_2 \leq \Re\lambda_1 \leq0$ is maintained across the whole relevant parameter domain.  Thus, not only does the coarse grid model $\dot u_j =\cop L {\bar u}_j =\op K_1 {\bar u}_j$
accurately model the fine grid dynamics,  the coarse grid model is the \emph{slowest} dynamics of the fine grid advection-dispersion.

\emph{Theorem~\ref{lem:cm} ensures an atttractive slow manifold exists in some neighbourhood of coupling $\gammb=0$\,; the spectrum~\eqref{eq:spec} demonstrates that the neighbourhood extends to include the case of fully coupled elements, $\gammb=1$\,.}

We usually cannot construct slow manifolds exactly, as done above; instead we usually approximate slow manifold by a multivariate power series.  Thus the practical issue is not just whether a slow manifold exists, but how well a truncated power series approximates the slow manifold.
Elementary algebra shows that a Taylor series of~\eqref{eq:spec} in~$\gammb$ converges at $\gammb=1$ provided 
\begin{displaymath}
\big[(1-\cc^2)^2\sin^2k+4\cc^2\big]\sin^2k<1\,,
\end{displaymath}
where $\cc=c/(2d)$ measures the advection relative to the dispersion.  For all parameter~$\cc$ there is a finite range of small wavenumbers~$k$ satisfying this inequality.   This argument leads to the following lemma.

\begin{lemma}
Finite truncations of the Taylor series of the slow operator eigenvalue~$\op K_1$ provide accurate approximations of the evolution of the coarse grid variables provided the solutions vary slowly enough across the grid.
\end{lemma}

\subsection{A normal form projects initial conditions}
\label{sec:nf}

Suppose we know the fine grid values~$u_j(0)$ at the initial time
$t=0$\,.  This subsection addresses the question: what coarse grid values should we give to~${\bar u}_j(0)$ for the coarse grid model to make accurate long term predictions?  

The obvious answer is wrong~\cite[e.g.]{Roberts89b, Cox93b, Roberts01a}:  even though we define
${\bar u}_j(t)=u_{2j}(t)$\,, we nonetheless should \emph{not} set the initial
${\bar u}_j(0)=u_{2j}(0)$\,.  The reason is that the transient dynamics of the subgrid scale dynamics modifies the appropriate initial value for
${\bar u}_j(0)$; this modification is sometimes called `initial slip' in physics \cite[e.g.]{Grad63, Geigenmuller83}.  In this subsection, a normal form coordinate transform of the fine grid dynamics clearly displays the correct initial conditions for the coarse dynamics.

In this discussion, restrict attention on initial conditions to the fully coupled case of coupling $\gammb=1$\,.  This restriction simplifies by avoiding the complicating detail of having variable~$\gammb$, and it focusses on the physically relevant case of full interelement coupling.

Consider the spectral decomposition of the dynamics of the fine grid of all the elements. Transform the fine grid evolution to its coarse grid `normal form' of the spectral decomposition
\begin{eqnarray*}&&
    \vec v_j(t)=\vec e_{j,1}{\bar u}_j(t)
    +\vec e_{j,2}{\bar v}_j(t)
    +\vec e_{j,3}{\bar w}_j(t)
    \,,
    \\\text{where}&&
    \dot {\bar u}_j=\op K_1 {\bar u}_j\,,\quad
    \dot {\bar v}_j=\op K_2 {\bar v}_j\,,\quad
    \dot {\bar w}_j=\op K_3 {\bar w}_j\,,
\end{eqnarray*}
for the operators~$\op K_n$ in~\eqref{eq:opeval} and for intraelement structure operators
\begin{equation}
    \vec e_{j,2}=
    \begin{bmatrix}
        \rat12c- d \\0\\ \rat12c+ d
    \end{bmatrix}
    \qtq{and}
    \vec e_{j,n}=
    \begin{bmatrix}
        \rat12c(\Ed^{-1}-1)+ d(\Ed^{-1}+1)\\
        \op K_n+2 d \\
        \rat12c(\Ed-1)+ d(\Ed+1)
    \end{bmatrix}
    \label{eq:ieso}
\end{equation}
for $n=1,3$\,.  I do not record the two extreme components $v_{j,\pm2}$ in these~$\vec e_{j,n}$ as $v_{j,\pm2}$ are identical to~$v_{j\pm1,0}$ when fully coupled, $\gammb=1$\,.  Within each of the fully coupled elements, a formal expression for the complete evolution on the fine grid is thus
\begin{equation}
    \vec v_j(t)=\vec e_{j,1}\exp(t\op K_1){\bar u}_j(0)
    +\vec e_{j,2}\exp(t\op K_2){\bar v}_j(0)
    +\vec e_{j,3}\exp(t\op K_3){\bar w}_j(0)
    \,,
    \label{eq:fineevol}
\end{equation}
for some constants ${\bar u}_j(0)$, ${\bar v}_j(0)$~and~${\bar w}_j(0)$.  For example,
from~\eqref{eq:opeval}, when advection $c=0$ the intraelement structure operators simplify to
\begin{displaymath}
    \vec e_{j,2}\propto
    \begin{bmatrix}
        -1 \\0\\ 1
    \end{bmatrix}
    \qtq{and}
    \vec e_{j,n}\propto 
    \begin{bmatrix}
        \rat12(\Ed^{-1}+1)\\
        \pm \bar\mu \\
        \rat12(\Ed+1)
    \end{bmatrix}
    \approx 
    \begin{bmatrix}
        1\\
        \pm 1 \\
        1
    \end{bmatrix}
\end{displaymath}
where this last approximate equality holds for fields varying slowly enough along the grids.  Thus $\vec e_{j,1}\approx (1,1,1)$ represents  the smoothest variations within each element, whereas $\vec e_{j,2}\approx (-1,0,1)$ and $\vec e_{j,3}\approx (1,-1,1)$ represents fine grid scale fluctuations within an element.\footnote{In essence, the element eigenvectors$\vec e_{j,n}$ are the natural wavelets~\cite[e.g.]{Daubechies92} for the advection dispersion dynamics.  The difference here is that, being adapted to the dynamics, the precise shape of the element eigenvectors depends upon the coupling with the neighbouring elements.}  Since these fine grid scale fluctuations decay rapidly in time~$t$, the long term slow dynamics on the slow manifold is just the restriction of~\eqref{eq:fineevol} to ${\bar v}_j={\bar w}_j=0$\,, namely
\begin{equation}
    \vec v_j(t)=\vec e_{j,1}\exp(t\op K_1t){\bar u}_j(0)\,.
    \label{eq:ltsd}
\end{equation}
We must choose the initial condition,~${\bar u}_j(0)$, for the coarse grid values so that this evolution exponentially quickly equals the fine grid dynamics $u_j(t)=\exp(t\op L )u_j(0)$ \emph{from the specified initial condition}.  Such a choice for the initial coarse grid value~${\bar u}_j(0)$ then realises the theoretical promise by Theorem~\ref{lem:cm}-\ref{it:cm2} of long term fidelity between coarse grid model and fine grid dynamics.  Elementary linear algebra determines the coarse grid values~${\bar u}_j(0)$ through evaluating the general solution~\eqref{eq:fineevol} at time $t=0$\,,
\begin{equation}
    \vec v_j(0)=\vec e_{j,1}{\bar u}_j(0)
    +\vec e_{j,2}{\bar v}_j(0)
    +\vec e_{j,3}{\bar w}_j(0)
    \,,
    \label{eq:init}
\end{equation}
and then take the inner product with the left eigenvector
\begin{displaymath}
    \vec z_{j,1}=
    \begin{bmatrix}
        \rat12c+ d \\[1ex]
        \op K_1 +2 d \\[1ex]
        -\rat12c+ d
    \end{bmatrix},
\end{displaymath}
to deduce the following lemma.

\begin{lemma}
For linear advection-dispersion, the initial coarse grid values are
\begin{equation}
    {\bar u}_j(0)=\frac{\vec z_{j,1}\cdot\vec v_j(0)}{\vec z_{j,1}\cdot\vec
    e_{j,1}}\,,
    \label{eq:proj}
\end{equation}
in terms of specified fine element values~$\vec v_j(0)$.  
\end{lemma}
Despite the definition that the coarse grid values
${\bar u}_j(t)=u_{2j}(t)$\,, the normal form coordinate transform accounts for dynamics in fast time initial transients so that the correct initial conditions for the coarse grid model is the nonlocal and weighted projection~\eqref{eq:proj}.

The initial condition mapping~\eqref{eq:proj} relates to multigrid iteration.  When advection $c=0$
\begin{displaymath}
    \vec z_{j,1}\propto
    \begin{bmatrix}
        1 \\
        2\bar\mu \\
        1
    \end{bmatrix}\approx
    \begin{bmatrix}
        1\\2\\1
    \end{bmatrix},    
\end{displaymath}
and in the case of slowly varying grid values, this projection from the fine grid initial values~$\vec v_j(0)$ to the coarse grid initial values is the classic multigrid restriction operator~\cite[e.g.]{Briggs01}: namely, that the coarse grid value is the average of the nearest fine grid values with a weighting of
$1:2:1$\,.

\paragraph{Uniquely prescribe fine element values}
We have an additional complication: on the fine grid, the odd grid values~$u_{2j\pm1}$ are shared between two neighbouring elements.  The grid value $u_{2j+1}$ is represented as both
$v_{j,1}$~and~$v_{j+1,-1}$, and both of these variables are treated as separate independent variables in the dynamics on each element.  We must resolve this separation.

Two independent suggestions resolve the separation with the same result.
My first suggestion to avoid conflict between the values of
$v_{j,1}$~and~$v_{j+1,-1}$ is to require that $v_{j,1}=v_{j+1,-1}$ at the initial time.
The shift operators rewrite this identity as
\begin{equation}
    E v_{j,0}=\Ed E^{-1} v_{j,0}\,,
    \label{eq:comp}
\end{equation}
where the coarse grid shift~$\Ed$ operates on the coarse grid, first subscript of~$v_{j,i}$, whereas the fine grid shift operator~$E$
operates on the fine grid, second subscript of~$v_{j,i}$.  For a domain with $m$~fine grid points, that is, $m/2$~coarse grid elements, the compatibility condition~\eqref{eq:comp} provides an additional
$m/2$~constraints to determine uniquely the $3m/2$~initial values~$v_{j,i}(0)$ within the elements from the $m$~fine grid values~$u_j(0)$.  My second suggestion is to choose $v_{j,\pm1}(0)$ so that the unphysical intermediate mode vanishes in the solution~\eqref{eq:fineevol}, that is, so that ${\bar v}_j(0)=0$ in the solution~\eqref{eq:fineevol}.  Then there will be no intermediate scale dynamics~$\exp(-2 d t)$ and the approach to the slow manifold model will be the quickest: the only rapidly decaying mode will be the $\vec e_{j,3}\exp(\op K_3t){\bar w}_j(0)$
mode which, from the spectrum~\eqref{eq:spec}, decays more rapidly than~$\exp(-2 d t)$.  Now relate ${\bar v}_j(0)$ directly to $\vec v_j(0)$ by multiplying~\eqref{eq:init} by the left eigenvector corresponding to
$\op K_2$ namely
\begin{eqnarray*}&&
    \vec z_{j,2}=
    \begin{bmatrix}
        \rat12c(\Ed-1)- d(\Ed+1) \\
        0\\
        \rat12c(\Ed^{-1}-1)+ d(\Ed^{-1}+1)
    \end{bmatrix}.
\end{eqnarray*}
Thus, noting $v_{j,\pm1}=E^{\pm1}v_{j,0}$\,,
\begin{eqnarray*}
    {\bar v}_j(0)&\propto&
    \left\{ \left[\rat12c(\Ed-1)- d(\Ed+1)\right]E^{-1} 
    +\left[\rat12c(\Ed^{-1}-1)+ d(\Ed^{-1}+1)\right]E
    \right\}v_{j,0}
    \\&=&
    \left\{ (\rat12c- d)(\Ed E^{-1}-E)
    +(\rat12c+ d)(\Ed^{-1}E-E^{-1})
    \right\}v_{j,0}\,.
\end{eqnarray*}
Consequently, ensure the mode $\exp(-2 d t)$ does not appear at all,
${\bar v}_j(0)=0$\,, by requiring $(\Ed E^{-1}-E)v_{j,0}=0$ which is precisely~\eqref{eq:comp}, and by requiring
$(\Ed^{-1}E-E^{-1})v_{j,0}=0$ which is again~\eqref{eq:comp} but just shifted to the left by the multiplication by the coarse grid shift~$\Ed^{-1}$.  Thus the condition~\eqref{eq:comp} ensures that neighbouring elements agree at their common points \emph{and} that the slow manifold, long term model is approached quickest.

\begin{lemma}
Choosing the embedding to $3m/2$-dimensions to satisfy~\eqref{eq:comp} at the initial time ensures that \eqref{eq:comp} is satisfied for all time in the linear, advection-dispersion dynamics on the fully coupled finite elements.
\end{lemma}

\subsection{Extend elements for a multigrid hierarchy}
\label{sec:eemh}

As discussed briefly in the Introduction, we aim to transform dynamics across a wide range of space-time scales using the multigrid hierarchy illustrated in Figure~\ref{fig:hsg}.  The Introduction used a model of~$\Ord{\gammb^2}$, see~\S\ref{sec:caasm}, to transform advection-dispersion on a fine grid to advection-dispersion \emph{of the same form} on a coarser grid.  This transform iterates simply across all scales.  However, when we seek more accuracy, say errors~$\Ord{\gammb^n}$ for $n>2$\,, the linear advection-dispersion dynamics~\eqref{eq:ad} transforms into a model of the form $d{\bar u}_j/dt=G(\bar u_{j-n+1},\ldots,\bar u_{j+n-1},\gammb)$ that involves $2(n-1)$~neighbouring coarse grid values.  For example,  to errors~$\Ord{\gammb^4}$, fine scale isotropic dispersion (equation~\eqref{eq:ad} with $c=f_j=0$) transforms to the coarser scale dispersion~\cite{Roberts07i}
\begin{equation}
    \frac{d{\bar u}_j}{dt}=
    d\left[\rat14\gammb\ded^2 -\rat1{64}\gammb^2\ded^4 
    +\rat1{512}\gammb^3\ded^6 \right] {\bar u}_j +\Ord{\gammb^4}\,,
    \label{eq:hig3}
\end{equation}
that through $\ded^4$ and~$\ded^6$ involves $\bar u_{j\pm2}$ and~$\bar u_{j\pm 3}$\,.
Consequently, to empower us to transform coarse models over a hierarchy of grids we must widen the elements defined in Figure~\ref{fig:explod} to include more fine grid points.  This subsection widens the elements while maintaining the spectrum~\eqref{eq:fineev} ensuring the centre manifold support~\cite[e.g.]{Carr81, Murdock03}.

This subsection, as seen in equation~\eqref{eq:hig3}, avoids the overdots for time derivatives as we invoke different time scales on each level of the hierarchy.  Interestingly, it eventuates that not only do we overlap the elements, but also,  in some sense, overlap the time scales.

\paragraph{The general form of linear dynamics on a grid}  Suppose at some level of the multigrid hierarchy we know the discrete operator governing the evolution of grid values~$u_j(t)$.  Decompose the discrete evolution operator as the sum
\begin{equation}
\frac{du_j}{dt}=\left[\op L_1+\op L_2+\op L_3+\cdots+\op L_{n-1}\right]u_j \,,
\label{eq:eemhb}
\end{equation}
where the $k$th~discrete operator~$\op L_k$ has stencil width~$2k+1$\,; that is, $\op L_ku_j$ only involves $u_{j-k},\ldots,u_{j+k}$\,.  This decomposition terminates, as written in~\eqref{eq:eemhb}, when we restrict attention, by working to errors~$\Ord{\gamma^n}$, to operators of some maximum finite width.    The decomposition is not unique as specified so far; however, as apparent in~\eqref{eq:hig3}, a specific unique decomposition naturally arises when we generate the models to errors~$\Ord{\gamma^n}$ in some coupling parameter~$\gamma$. Thus suppose there is a natural `ordering' parameter~$\gamma$ such that, instead of~\eqref{eq:eemhb}, the discrete evolution equation may be written
\begin{equation}
\frac{du_j}{dt}=\left[\op L_1+\gamma\op L_2+\gamma^2\op L_3+\cdots+\gamma^{n-2}\op L_{n-1}\right]u_j \,.
\label{eq:gadn}
\end{equation}
At all levels, except the very finest level~$0$, this natural parameter~$\gamma$ is to be the coupling parameter of the finite elements of the grid one level finer than than current level.  As always, we suppose that evaluation of~\eqref{eq:gadn} at $\gamma=1$ gives the physically relevant model~\eqref{eq:eemhb}, whereas $\gamma=0$ provides a base for theory to support models at non-zero~$\gamma$.  Additionally insisting on the operator~$\op L_1$ being conservative implies $\op L_1$~must represent advection-dispersion dynamics and implies that the $\gamma=0$ dynamics, $d u_j/dt=\op L_1u_j$\,, provides the same sound base for applying centre manifold theory.   Note that the coupling parameter of finite elements at the current level is still~$\gammb$.  That is, still couple neighbouring elements with the condition~\eqref{eq:ibc}.

Anticipating the support by centre manifold theory, derived in a couple of paragraphs, we expect to construct a coarse grid model of the dynamics~\eqref{eq:gadn} in the form
\begin{equation}
\frac{d\bar u_j}{dt}=\left[ \gammb\cop L_1 +\gammb^2\cop L_2 +\gammb^3\cop L_3 +\cdots +\gammb^{n-1}\cop L_{n-1}\right]\bar u_j
+\Ord{\gammb^n,\gamma^{n-1}}\,,
\label{eq:cgm}
\end{equation}
for some coarse grid operators~$\cop L_k$ (implicitly a function of the artificial~$\gamma$) which will be of stencil width~$2k+1$ as the parameter~$\gammb$ counts the number of interelement communications.  The renormalising transformation requires two extra ingredients: first remove the fine grid ordering by setting $\gamma=1$ (so operators~$\cop L_k$ are no longer a function of~$\gamma$); and second introduce a coarse grid time scale $\bar t=t/\gammb$ (which is the same time when $\gammb=1$), then, upon dividing by~$\gammb$, the coarse grid dynamics become
\begin{equation}
\frac{d\bar u_j}{d\bar t}=\left[\cop L_1+\gammb\cop L_2+\gammb^2\cop L_3+\cdots+\gammb^{n-2}\cop L_{n-1}\right]\bar u_j +\Ord{\gammb^{n-1}}\,.
\label{eq:cgmn}
\end{equation}
The coarse model~\eqref{eq:cgmn} has exactly the same form as the fine model~\eqref{eq:gadn}.  By introducing the coupling~\eqref{eq:ibc} across all levels of the hierarchy, and by introducing a hierarchy of times, which all collapse to the same real time when $\gammb=1$\,, and working to some order of error in coupling, models of the form~\eqref{eq:gadn} are transformed and renormalised across the entire multigrid hierarchy.

\paragraph{Widen the elements}
Assume we wish to construct slow manifolds to errors~$\Ord{\gammb^n,\gamma^{n-1}}$ with the aim of using centre manifold theory to support the modelling of~\eqref{eq:gadn} by~\eqref{eq:cgmn}.  Extend Section~\ref{sec:two} by widening the $j$th~element to extend over the interval~$[x_{j-n},x_{j+n}]$ and also to possess the $(2n+1)$~fine grid variables $\vec v_j=(v_{j,-n},\ldots,v_{j,n})$\,.  These extra variables are not extra degrees of freedom. Let these fine grid variables evolve according to
\begin{equation}
\frac{dv_{j,i}}{dt}=\op L_1v_{j,i} +\gamma\op L_2v_{j,i} +\gamma^2\op L_3v_{j,i} +\cdots+\gamma^{n-2}\op L_{n-1}v_{j,i} \,, \quad |i|<n\,,
\label{eq:gadnv}
\end{equation}
where we adopt the unusual convention that when applied within the elements, the operator $\op L_k v_{j,i}$ is its original definition when $|i|+k\leq n$ but is \emph{zero otherwise} (for $|i|+k>n$).  Adopting this convention ensures that the operators on the right-hand side of~\eqref{eq:gadnv} do not `poke outside' of the $j$th~element; in effect, this convention truncates the sum in~\eqref{eq:gadnv} to remain within the $j$th~element.  Such truncation incurs an error~$\Ord{\gamma^{n+1-|i|}}$ in the evolution of a variable~$v_{j,i}$.  However, as variable~$v_{j,i}$ only affects the crucial central core variables of the element, $v_{j,i'}$~for $|i'|\leq 2$ as shown in Figure~\ref{fig:explod}, via terms of~$\Ord{\gamma^{|i|-2}}$, the net effect of this conventional truncation is an error~$\Ord{\gamma^{n-1}}$ which is the same as the assumed order of error of the analysis.  The coupling condition~\eqref{eq:ibc} closes the dynamics on these widened elements.  In essence we do not have new dynamics outside of the central core of each element, instead, in effect, we simply extrapolate the dynamics to the outside of the central core.

\paragraph{Centre manifold theory support}  When the fine grid `ordering' parameter $\gamma=0$ and interelement coupling parameter $\gammb=0$ the dynamics on the $m/2$~elements reduces to
\begin{equation}
\frac{dv_{j,i}}{dt}=\op L_1v_{j,i}
\quad\text{for }|i|<n\,,
\qtq{and} v_{j,\pm2}=v_{j,0}\,.
\label{eq:m0dyn}
\end{equation}
Each element is decoupled from the others.  The general conservative, linear, operator is the advection-dispersion operator,  $\op L_1=-c\mud+d\delta^2$ for some constants $c$~and~$d$.  As for the earlier~\eqref{eq:fined}, for each of the extended elements there are still precisely three eigenvalues of~\eqref{eq:m0dyn}, namely $\lambda=0,-2d,-4d$\,.  Corresponding eigenvectors are the constant $v_{j,i}\propto1$\,, the artificial $v_{j,i}\propto\sin(i\pi/2)\left[({1+\cc})/({1-\cc})\right]^{i/2}$, and the zig-zag mode $v_{j,i}\propto(-1)^i$\,.  Consequently, centre manifold theory implies Theorem~\ref{lem:cm} also applies to the system~\eqref{eq:gadnv} with coupling conditions~\eqref{eq:ibc} to ensure: firstly, that an $m/2$~dimensional slow manifold exists for the dynamics of the coupled elements; secondly, the coarse scale dynamics on the slow manifold are attractive; and thirdly, that we may construct the slow manifold to any desired error---this section assumes errors~$\Ord{\gamma^{n-1},\gammb^n}$.  

The next subsection proceeds to briefly explore the resultant models of advection and dispersion over a hierarchy of multiscale grids as supported by this theory.

\section{Multiscale modelling iterates transformations}
\label{sec:multi}

This section explores three example applications of transforming dynamics repeatedly across the wide range of length and time scales on a multigrid hierarchy.  Section~\ref{sec:defd} shows how continuum diffusion emerges from microscale dispersion. Section~\ref{sec:rad} deonstrates that the nonlinear transformation from one scale to another of linear advection-dispersion has a fixed point of a stable upwind model. Section~\ref{sec:anbd} discusses briefly the transformation of the nonlinear Burgers' \pde.

\subsection{Diffusion emerges from discrete dispersion}
\label{sec:defd}

The multiscale modelling of discrete dispersion, when the advection coefficient $c=0$\,, reduces to a remarkably simple linear transformation.  Here we explore the exact slow manifold transformation from a fine grid to a coarser grid.
Iterating this transformation proves that, in the absence of advection, the continuum diffusion equation naturally emerges very quickly on macroscales.  

Linear dynamics which are left-right symmetric (isotropic) can be expressed in terms of only even order central differences.  Our slow manifold, multiscale modelling preserves this form.  Suppose the evolution at grid level~$\ell$ is governed by
\begin{equation}
    \frac{du_j\g\ell}{dt\g\ell} 
    =\sum_{p=1}^\infty \gamma^{p-1}c_{p}\g\ell {\delta\g\ell}^{2p}u_j\g\ell \,,
    \label{eq:gn}
\end{equation}
for some coefficients~$c_{p}$\,; for example, the second order difference coefficient $c_1=d$ used earlier.  In practical constructions, invoking an error~$\Ord{\gamma^{n-1}}$ truncates to a finite sum this `in principle' infinite sum.  Computer algebra~\cite{Roberts07i}, supported by the theory of Section~\ref{sec:eemh}, derives the dynamics at the next coarser level of the multiscale hierarchy, namely
\begin{eqnarray}
    \frac{du_j\g{\ell+1}}{dt\g{\ell+1}}&=& \big[
    \rat14c_1\g{\ell}{\delta\g{\ell+1}}^2
    \nonumber\\&&{}
    +\gammb\Big(-\rat1{64}c_1\g{\ell} +\rat1{16}c_2\g{\ell} \Big){\delta\g{\ell+1}}^4
    \nonumber\\&&{}
    +\gammb^2\Big(\rat1{512}c_1\g{\ell} -\rat1{128}c_2\g{\ell}
    +\rat1{64}c_3\g{\ell} \Big){\delta\g{\ell+1}}^6
    \nonumber\\&&{}
    +\gammb^3\Big(-\rat5{16384}c_1\g{\ell} +\rat5{4096}c_2\g{\ell}
    -\rat3{1024}c_3\g{\ell} +\rat1{256}c_4\g{\ell} \Big){\delta\g{\ell+1}}^8
    \nonumber\\&&{}
    \big]u_j\g{\ell+1}+\Ord{\gammb^4}\,. \label{eq:dnlmh}
\end{eqnarray}
That is, to model level~$\ell$ dynamics at level~$(\ell+1)$ the coefficients in~\eqref{eq:gn} transform according to the linear transform 
\begin{equation}
    \vec c\g{\ell+1}=\cT\vec c\g{\ell}
    \qtq{where}
    \cT=
    \begin{bmatrix}
        \rat14 & 0 & 0 & 0 & \cdots \\[1ex]
        -\rat1{64}&\rat1{16} & 0 & 0 & \cdots \\[1ex]
        +\rat1{512}&-\rat1{128}&\rat1{64}&  0 & \cdots \\[1ex]
        -\rat5{16384}&\rat5{4096}&-\rat3{1034}&\rat1{256} &\\
        \vdots&\vdots&\vdots& &\ddots
    \end{bmatrix}.
    \label{eq:ct}
\end{equation}
By induction, the level~$\ell$ dynamics have centred difference coefficients
$\vec c\g\ell =\cT^\ell\vec c\g0$\,.  Consequently, the dynamics that emerge on macroscale grids are determined by the powers~$\cT^\ell$ for large~$\ell$.
Since~$\cT$ is triangular the powers are  simple: the dominant structure for large~$\ell$ corresponds to that of the leading eigenvalue~$1/4$; its eigenvector gives the centred difference coefficients that emerge on the macroscale as
\begin{equation}
    \vec c\g\ell \sim \frac{c_1\g0}{4^\ell}\big( 1, -\rat1{12}, \rat1{90},
    -\rat1{560}, \ldots)
    \quad\text{as }\ell\to\infty\,.
\end{equation}
Recognise in this vector the coefficients of various powers of centred differences in a discrete representation of the continuum diffusion operator~$\partial^2/\partial x^2$.  That is,\footnote{This emergence of diffusion is proved here only up to the order of error in the truncation used here.}
\emph{continuum diffusion emerges on the macroscale for all isotropic,
conservative, linear, continuous time, microscale dynamics provided there is some component of~$\delta^2$ in the microscale
($c_1\g0\neq 0$).}

What is novel here?  That continuum diffusion emerges on macroscopic scales has been well known for centuries.  The novelty is the centre manifold theory framework I set up to prove this well known fact.  This framework illuminates issues and empowers us to analyse much more difficult nonlinear dynamics, Section~\ref{sec:anbd}, and potentially stochastic problems.

Furthermore, the framework shows that a consistent truncation in the interelement coupling parameter~$\gammb$ generates a macroscopic approximation to continuum diffusion that is of the same order of error in~$\gammb$.  (For example, in the Introduction we discussed multiscale modelling with truncations to~$\Ord{\gammb^2}$.)  This consistency follows because truncating the mapping operator~$\cT$, given in~\eqref{eq:ct}, simply truncates its spectrum \emph{and} truncates its eigenvectors (as $\cT$~is triangular).

\subsection{Renormalise advection-dispersion}
\label{sec:rad}

Now consider advection-dispersion on a multiscale hierarchy.  Although the dynamics are linear, and in contrast to the previous subsection, the transformation from one level to another in the hierarchy is nonlinear.  

For example, suppose the microscopic dynamics is simply the discrete $\dot u_j=[-\mu\delta+\delta^2]u_j$\,.  Then computer algebra~\cite{Roberts07i} derives that the multigrid hierarchy of dynamic models is
\begin{eqnarray}
\frac{du_j\g0}{dt\g0}&=&[-\mu\delta+\delta^2]u_j\g0 \,,
\label{eq:rad0}\\
\frac{du_j\g1}{dt\g1}&=&\rat12\big[-\mu\delta +\rat5{8}\delta^2
+\gamma(-\rat1{8}\delta^2 +\rat5{32}\mu\delta^3 -\rat{41}{512}\delta^4 )
\big]u_j\g1 +\Ord{\gamma^2} \,,\\
\frac{du_j\g2}{dt\g2}&=&\rat14\big[-\mu\delta +\rat{41}{80}\delta^2
+\gamma(-\rat{21}{80}\delta^2 +\rat{861}{3200}\mu\delta^3 
-\rat{68901}{512000}\delta^4 )
\big]u_j\g2 
\nonumber\\&&{}
+\Ord{\gamma^2} \,,\\
\frac{du_j\g3}{dt\g3}&=&\rat18\big[-\mu\delta +0.50015\delta^2
\nonumber\\&&{}
+\gamma(-0.37515\delta^2 +0.37527\mu\delta^3 
-0.18763\delta^4 )
\big]u_j\g3 +\Ord{\gamma^2} \,,\\
&\vdots&\nonumber\\
\frac{du_j\g9}{dt\g9}&=&\rat1{2^9}\big[-\mu\delta +0.50000\delta^2
\nonumber\\&&{}
+\gamma(-0.49805\delta^2 +0.49805\mu\delta^3 
-0.24902\delta^4 )
\big]u_j\g9 +\Ord{\gamma^2} \,,\quad
\label{eq:rad9}
\end{eqnarray}
where for simplicity I omit the level of the discrete mean and difference operators. Evidently, as the level~$\ell$ increases, and upon renormalising the time scale by the factor of~$2^\ell$ (the grid step), these models approach a fixed point corresponding to an upwind model of the advection.  As is well known, advection dominates diffusion on large scales.  This centre manifold supported multiscale transformation preserves the advection speed, and does so with stable upwind differencing.

Now explore the general mapping of linear conservative dynamics from one level on the multigrid to another. Generalise the form~\eqref{eq:gn} for isotropic dynamics to the general finite difference representation of conservative linear operators:
\begin{equation}
    \frac{du_j }{dt } 
    =\sum_{p=1}^\infty \gamma^{p-1}\left\{
    \sum_{k=1}^{p}\left(
    c_{p,2k-1}\mu{\delta}^{2k-1}
    +c_{p,2k} {\delta }^{2k}\right)
    \right\}u_j \,,
    \label{eq:gnad}
\end{equation}
for some coefficients $c_{p,k}$ where $c_{1,1}=-c$ and $c_{1,2}=d$ as used earlier.  The operator in braces~$\{\,\}$, called $\op L_p$~earlier, represents a general operator of stencil width~$2p+1$.  For example, working to error~$\Ord{\gamma^3,\gammb^2}$, computer algebra~\cite{Roberts07i} derives the model at the next level of the the multiscale hierarchy to be, in gory detail,
\begin{eqnarray}
\frac{d\bar u_j}{d\bar t} &=&
\left\{\rat12\big[c_{1,1}+c_{2,1}\big]\mud
+\left[ \frac14(c_{1,2}+c_{2,2}) 
\right.\right.\nonumber\\&&\left.\left.\quad{}
+\frac{c_{1,1}}{16c_{1,2}}(c_{1,1} +2c_{2,1} -4c_{2,3})
+\frac{c_{1,1}^2}{16c_{1,2}^2}(-c_{2,2}+4c_{2,4})\right]\ded^2
\right\}\bar u_j
\nonumber\\&&{}+\gammb\left\{ 
\left[ \frac{c_{1,1}}{16c_{1,2}}(-c_{1,1}-2c_{2,1}+4c_{2,3}) 
+\frac{c_{1,1}^2}{16c_{1,2}^2}(c_{2,2}-4c_{2,4}) \right]\ded^2
\right.\nonumber\\&&\left.\quad{}
+\left[\frac1{16}(-c_{1,1}-c_{2,1}-2c_{2,3})
+\frac{c_{1,1}^2}{64c_{1,2}^2}(-c_{1,1}-3c_{2,1}+3c_{2,3})
\right.\right.\nonumber\\&&\left.\left.\qquad{}
+\frac{c_{1,1}^3}{32c_{1,2}^3}(c_{2,2}-4c_{2,4}) \right]\mud^3
\right.\nonumber\\&&\left.\quad{}
+\left[\frac1{64}(-c_{1,2}-c_{2,2}-4c_{2,4})
+\frac{3c_{1,1}}{128c_{1,2}}(-c_{1,1}-c_{2,1}+c_{2,3})
\right.\right.\nonumber\\&&\left.\left.\qquad{}
+\frac{3c_{1,1}^2}{128c_{1,2}^2}(c_{2,2}-c_{2,4})
+\frac{c_{1,1}^3}{1024c_{1,2}^3}(-c_{1,1}-4c_{2,1}+8c_{2,3})
\right.\right.\nonumber\\&&\left.\left.\qquad{}
+\frac{3c_{1,1}^4}{1024c_{1,2}^4}(c_{2,2}-4c_{2,4})\right]\ded^4
\right\} \bar u_j
\nonumber\\&&{}+\Ord{\gammb^2}\,.
\label{eq:gnpad}
\end{eqnarray}
This general mapping from~\eqref{eq:gnad} to~\eqref{eq:gnpad} governs the particular exapmple hierarchy of models \eqref{eq:rad0}--\eqref{eq:rad9}.  Fine grid scale interactions generate the nonlinear dependence upon coefficients shown in the transformation to~\eqref{eq:gnpad}.  The example hierarchy \eqref{eq:rad0}--\eqref{eq:rad9} shows that when we scale time by a further factor of two in each transformation, to correspond to the time scale of advection of a grid of twice the spacing, then the multiscale transformation possess a fixed point.   Returning to the general transformation from~\eqref{eq:gnad} to~\eqref{eq:gnpad}, rescaling time by a factor of two, computer algebra finds precisely two non-trivial fixed points of the multiscale transformation:
\begin{eqnarray*}
\frac{du_j\g*}{dt\g*}&=&c_*\left\{\mp\mu\delta +\rat12\delta^2
+\gamma\big[-\rat12\delta^2 \pm\rat12\mu\delta^3 
-\rat14\delta^4 \big]
\right\}u_j\g* +\Ord{\gamma^2} \,,
\end{eqnarray*}
for some speed~$c_*$ (positive) which will depend upon the precise microscopic system.  These fixed points are purely upwind macroscale models of the advection and dispersion dynamics.  Such stable upwind models naturally emerge from our rational transformation of dynamics based upon dynamical systems theory.

\subsection{Approximate the nonlinear Burgers' dynamics}
\label{sec:anbd}

Burgers' partial differential equation, $\partial u/\partial t+u\,\partial u/\partial x=\partial^2 u/\partial x^2$\,, is frequently invoked as a benchmark problem in nonlinear spatio-temporal dynamics as it involves the important physical mechanisms of dissipative diffusion and nonlinear advection.  As an example of a nonlinear application of our multiscale methodology, suppose Burgers' \pde\ is spatially discretised to
\begin{equation}
\frac{du_j\g0}{dt\g0} = \delta^2u_j\g0 
-\alpha u_j\g0\mu\delta u_j\g0\,,
\label{eq:burg0}
\end{equation} 
where the time scale~$t\g0$ is chosen to make the coefficient unity for the centred difference approximation~$\delta^2u_j$ of the diffusion~$\partial^2u/\partial x^2$.  Take equation~\eqref{eq:burg0} to be the microscale discrete nonlinear dynamics. The parameter~$\alpha$ measures the importance of the nonlinear advection on this microscopic scale.  Section~\ref{sec:two} places the coarse grid modelling of such nonlinear discrete dynamics within the purview of centre manifold theory.

For relatively small parameter~$\alpha$, straightforward modifications of the computer algebra for the earlier linear dynamics~\cite{Roberts07i} analyses nonlinear problems.  The reason is that as long as the nonlinearity is relatively weak, small~$\alpha$, the dominant mechanism in each element is the linear dissipation of~$\delta^2$ just as for the linear dynamics.  Our multiscale modelling transforms the fine grid dynamics~\eqref{eq:burg0} into the level one dynamics~\eqref{eq:burg1}; applying the multiscale modelling again transforms the level one dynamics~\eqref{eq:burg1} into the level two dynamics~\eqref{eq:burg2}.
\begin{eqnarray}
\frac{du_j\g1}{dt\g1}&=&
\left[\frac14 +\frac1{16}\alpha^2{u_j\g1}^2\right]\delta^2u_j\g1 
-\frac\gamma{64}\delta^4u_j\g1
-\frac\alpha2u_j\g1\delta^2u_j\g1
\nonumber\\&&{}
+\frac{\alpha\gamma}{64}\left[4u_j\g1\mu\delta^3u_j\g1 +(\delta^2u_j\g1)\mu\delta^3u_j\g1 +(\delta^4u_j\g1)\mu\delta u_j\g1 \right]
\nonumber\\&&{}
+\Ord{\gamma^2+\alpha^3} \,, \label{eq:burg1} \\
\frac{du_j\g2}{dt\g2}&=&
\left[\frac1{16} +\frac5{64}\alpha^2{u_j\g2}^2\right]\delta^2u_j\g2 
-\frac{5\gamma}{1024}\delta^4u_j\g2
-\frac\alpha4u_j\g2\delta^2u_j\g2
\nonumber\\&&{}
+\frac{5\alpha\gamma}{512}\left[4u_j\g2\mu\delta^3u_j\g2 +(\delta^2u_j\g2)\mu\delta^3u_j\g2 +(\delta^4u_j\g2)\mu\delta u_j\g2 \right]
\nonumber\\&&{}
+\Ord{\gamma^2+\alpha^3} \,. \label{eq:burg2}
\end{eqnarray} 
Evidently we could continue this transformation across many more levels of a multigrid hierarchy.

The transformation from~\eqref{eq:burg0} to~\eqref{eq:burg1}, to~\eqref{eq:burg2} is based upon small nonlinearity~$\alpha$.  However, as we should expect, the nonlinear advection appears to become more important at larger scales: the relative magnitude of the nonlinear enhancement to dissipation,~$\alpha^2 \big(u_j\g\ell\big)^2$, increases when going from level~$0$ to level~$2$.  After transforming over enough levels, the nonlinearity will begin to dominate the linear basis of the analysis here; at that length scale I expect the discrete dynamics to morph into a qualitatively new form, one dominated by nonlinear advection.  Although these emergent dynamics cannot be captured by the transformation used here,  a generalisation of the algebra to being based about a nonlinear subspace of piecewise constant solutions may be feasible.  In that case, the centre manifold theory of Section~\ref{sec:two} would still support the multiscale modelling of the strongly nonlinear dynamics.

\section{Conclusion}

This article introduces a new dynamical systems approach to modelling and linking dynamics across a multigrid hierarchy.

Because we recover continuum diffusion,~\S\ref{sec:defd}, and upwind advection,~\S\ref{sec:rad}, on macroscales  we  are reassured that the process of modelling from one grid to the  next coarser grid is indeed sound, as claimed by centre manifold theory, \S\ref{sec:two}.  Further, errors do not appear to accumulate when we iterate the modelling transformation across many changes in length scales.

At all lengths scales in the hierarchy of models, centre manifold theory assures us that the model on each scale is exponentially attractive,~\S\ref{sec:csda}, and provides an estimate of the rate of attraction.  This theoretical support applies for the finite spectral gaps on the multigrid hierarchy.  

The geometric picture of invariant slow manifolds also provides a rationale for providing initial conditions for the  models at each length scale \cite[e.g.]{Roberts89b, Cox93b, Roberts01a}.  Section~\ref{sec:nf} connects appropriate initial conditions with the restriction projection of multigrid solution of linear equations~\cite[e.g.]{Briggs01}.
    
In addition to providing the dynamics at all length scales in the hierarchy, this approach also provides intraelement structures  realised by the dynamics of the grid values: in terms of the  level~$(\ell+1)$ dynamic variables, equation~\eqref{eq:ltsd}  with~\eqref{eq:ieso} describe the corresponding structures on the level~$\ell$ grid.  In some sense, equation~\eqref{eq:ieso} provide  `wavelets' for each grid scale \cite[e.g.]{Daubechies92,  Jaffard01}.  Our modelling connects the dynamics of wavelets across  a hierarchy of length scales.

All the analysis herein is for dynamics in one space dimension.  Just as for holistic discretisation of \pde{}s~\cite{MacKenzie05}, I expect extension to higher space dimensions will be straightforward.  This article focussed on dynamics which to a first approximation could be modelled by advection-dispersion equations; extension to dynamics of necessarily higher-order, such as a discrete Kuramoto-Sivashinsky equation, could also be analogous to the approach of holsitic discretisation~\cite{MacKenzie00a, MacKenzie05a}.   Similarly, extension to stochastic mutliscale dynamics could be analogous to that of holistic discretisation of stochastic dynamics~\cite[e.g.]{Roberts05c, Roberts06g}.

\bibliographystyle{plain}
\bibliography{ajr,bib}

\begin{thebibliography}{10}

\bibitem{Arnold03}
Ludwig Arnold.
\newblock {\em Random Dynamical Systems}.
\newblock Springer Monographs in Mathematics. Springer, June 2003.

\bibitem{Boxler89}
P.~Boxler.
\newblock A stochastic version of the centre manifold theorem.
\newblock {\em Probab.\ Th.\ Rel.\ Fields}, 83:509--545, 1989.

\bibitem{Brandt01}
Achi Brandt.
\newblock Multiscale scientific computation: review 2001.
\newblock In T.~F.~Chan T.~J.~Barth and R.~Haimes, editors, {\em Multiscale and
  Multiresolution Methods: Theory and Applications}, pages 1--96.
  Springer--Verlag, Heidelberg, 2001.

\bibitem{Brandt06}
Achi Brandt.
\newblock Methods of systematic upscaling.
\newblock Technical report, Department of Computer Science and Applied
  Mathematics, The Weizmann Institute of Science, March 2006.

\bibitem{Briggs01}
William~L. Briggs, Van~Emden Henson, and Steve~F. McCormick.
\newblock {\em A multigrid tutorial, second edition}.
\newblock SIAM, 2nd edition, 2001.

\bibitem{Carr81}
J.~Carr.
\newblock {\em Applications of centre manifold theory}, volume~35 of {\em
  Applied Math. Sci.}
\newblock Springer--Verlag, 1981.

\bibitem{Chao95}
Xu~Chao and A.~J. Roberts.
\newblock On the low-dimensional modelling of {Stratonovich} stochastic
  differential equations.
\newblock {\em Physica~A}, 225:62--80, 1996.
\newblock \doi{10.1016/0378-4371(95)00387-8}.

\bibitem{Cox93b}
S.~M. Cox and A.~J. Roberts.
\newblock Initial conditions for models of dynamical systems.
\newblock {\em Physica~D}, 85:126--141, 1995.
\newblock \doi{10.1016/0167-2789(94)00201-Z}.

\bibitem{Daubechies92}
Ingrid Daubechies.
\newblock {\em Ten lectures on wavelets}.
\newblock SIAM, 1992.

\bibitem{Dolbow04}
J.~Dolbow, M.~A. Khaleel, and J.~Mitchell.
\newblock Multiscale mathematics initiative: A~roadmap. {Report} from the 3rd
  {DoE} workshop on multiscale mathematics.
\newblock Technical report, Department of Energy, USA,
  \url{http://www.sc.doe.gov/ascr/mics/amr}, December 2004.

\bibitem{E04}
Weinan E, Bjorn Engquist, Xiantao Li, Weiqing Ren, and Eric Vanden-Eijnden.
\newblock The heterogeneous multiscale method: A review.
\newblock Technical report,
  \url{http:www.math.princeton.edu/multiscale/review.pdf}, 2004.

\bibitem{Elphick87a}
C.~Elphick, G.~Iooss, and E.~Tirapegui.
\newblock Normal form reductions for time-periodically driven differential
  equations.
\newblock {\em Phys. Lett.~A}, 120:459--463, 1987.

\bibitem{Geigenmuller83}
U.~Geigenm\"uller, U.~M. Titulaer, and B.~U. Felderhof.
\newblock Systematic elimination of fast variables in linear systems.
\newblock {\em Physica~A}, 119:41--52, 1983.

\bibitem{Grad63}
H.~Grad.
\newblock Asymptotic theory of the {Boltzmann} equation.
\newblock {\em Phys. Fluids}, 6:147--181, 1963.

\bibitem{Griebel06}
Michael Griebel, Daniel Oeltz, and Panayot Vassilevski.
\newblock Space-time approximation with sparse grids.
\newblock {\em SIAM Journal on Scientific Computing}, 28:701--727, 2006.

\bibitem{Jaffard01}
St\'ephane Jaffard, Yves Meyer, and Robert~D. Ryan.
\newblock {\em Wavelets: Tools for Science and Technology}.
\newblock SIAM, 2001.

\bibitem{MacKenzie00a}
T.~Mackenzie and A.~J. Roberts.
\newblock Holistic finite differences accurately model the dynamics of the
  {Kuramoto--Sivashinsky} equation.
\newblock {\em ANZIAM~J.}, 42(E):C918--C935, 2000.
\newblock \url{http://anziamj.austms.org.au/V42/CTAC99/Mack}.

\bibitem{MacKenzie05a}
T.~MacKenzie and A.~J. Roberts.
\newblock Accurately model the {Kuramoto--Sivashinsky} dynamics with holistic
  discretisation.
\newblock {\em SIAM J.~Applied Dynamical Systems}, 5(3):365--402, 2006.
\newblock \doi{10.1137/050627733}
  \url{http://epubs.siam.org/SIADS/volume-05/art_62773.html}.

\bibitem{MacKenzie05}
Tony MacKenzie.
\newblock {\em Create accurate numerical models of complex spatio-temporal
  dynamical systems with holistic discretisation}.
\newblock PhD thesis, University of Southern Queensland, 2005.

\bibitem{Murdock03}
James Murdock.
\newblock {\em Normal forms and unfoldings for local dynamical systems}.
\newblock Springer Monographs in Mathematics. Springer, 2003.

\bibitem{npl61}
{National Physical Laboratory}.
\newblock {\em Modern Computing Methods}, volume~16 of {\em Notes on Applied
  Science}.
\newblock Her Majesty's Stationery Office, London, 2nd edition, 1961.

\bibitem{Pavliotis06a}
G.~A. Pavliotis and A.~M. Stuart.
\newblock {\em An introduction to multiscale methods}.
\newblock University of Warwick and Imperial College London, 2006.
\newblock \url{ http://www.maths.warwick.ac.uk/~stuart}.

\bibitem{Pavliotis07}
G.~A. Pavliotis and A.~M. Stuart.
\newblock Multiscale methods: averaging and homogenization.
\newblock Technical report, \url{http://www.ma.ic.ac.uk/~pavl}, 2007.

\bibitem{Roberts89b}
A.~J. Roberts.
\newblock Appropriate initial conditions for asymptotic descriptions of the
  long term evolution of dynamical systems.
\newblock {\em J.~Austral. Math. Soc.~B}, 31:48--75, 1989.

\bibitem{Roberts96a}
A.~J. Roberts.
\newblock Low-dimensional modelling of dynamics via computer algebra.
\newblock {\em Computer Phys. Comm.}, 100:215--230, 1997.

\bibitem{Roberts98a}
A.~J. Roberts.
\newblock Holistic discretisation ensures fidelity to {Burgers'} equation.
\newblock {\em Applied Numerical Modelling}, 37:371--396, 2001.

\bibitem{Roberts01a}
A.~J. Roberts.
\newblock Holistic projection of initial conditions onto a finite difference
  approximation.
\newblock {\em Computer Phys. Comm.}, 142:316--321, 2001.

\bibitem{Roberts99d}
A.~J. Roberts.
\newblock Simple and fast multigrid solution of {Poisson's} equation using
  diagonally oriented grids.
\newblock {\em ANZIAM~J.}, 43(E):E1--E36, July 2001.
\newblock \url{http://anziamj.austms.org.au/V43/E025}.

\bibitem{Roberts00a}
A.~J. Roberts.
\newblock A holistic finite difference approach models linear dynamics
  consistently.
\newblock {\em Mathematics of Computation}, 72:247--262, 2002.
\newblock \url{http://www.ams.org/mcom/2003-72-241/S0025-5718-02-01448-5}.

\bibitem{Roberts03c}
A.~J. Roberts.
\newblock A step towards holistic discretisation of stochastic partial
  differential equations.
\newblock In Jagoda Crawford and A.~J. Roberts, editors, {\em Proc. of 11th
  Computational Techniques and Applications Conference CTAC-2003}, volume~45,
  pages C1--C15, December 2003.
\newblock [Online] \url {http://anziamj.austms.org.au/V45/CTAC2003/Robe}
  [December 14, 2003].

\bibitem{Roberts05c}
A.~J. Roberts.
\newblock Resolving the multitude of microscale interactions accurately models
  stochastic partial differential equations.
\newblock {\em LMS J.~Computation and Maths}, 9:193--221, 2006.
\newblock \url{http://www.lms.ac.uk/jcm/9/lms2005-032}.

\bibitem{Roberts07i}
A.~J. Roberts.
\newblock Computer algebra models dynamics on a multigrid across multiple
  length and time scales.
\newblock Technical report, University of Southern Queensland,
  \url{http://eprints.usq.edu.au/3373/}, November 2007.

\bibitem{Roberts06g}
A.~J. Roberts.
\newblock Subgrid and interelement interactions affect discretisations of
  stochastically forced diffusion.
\newblock In Wayne Read, Jay~W. Larson, and A.~J. Roberts, editors, {\em
  Proceedings of the 13th Biennial Computational Techniques and Applications
  Conference, CTAC-2006}, volume~48 of {\em ANZIAM~J.}, pages C168--C187.
  \url{http://anziamj.austms.org.au/ojs/index.php/ANZIAMJ/article/view/36},
  June 2007.

\bibitem{Roberts06k}
A.~J. Roberts.
\newblock Normal form transforms separate slow and fast modes in stochastic
  dynamical systems.
\newblock {\em Physica~A}, 387:12--38, 2008.

\bibitem{Samaey03b}
G.~Samaey, I.~G. Kevrekidis, and D.~Roose.
\newblock The gap-tooth scheme for homogenization problems.
\newblock {\em SIAM Multiscale Modeling and Simulation}, 4:278--306, 2005.
\newblock \doi{10.1137/030602046}.

\bibitem{Verhulst05}
Ferdinand Verhulst.
\newblock {\em Methods and applications of singular perturbations: boundary
  layers and multiple timescales}, volume~50 of {\em Texts in Applied Maths}.
\newblock Springer, 2005.

\end{thebibliography}

\end{document}